\documentclass[11pt,preprint]{aastex}
\pdfoutput=1
\usepackage{bm}
\usepackage{graphicx}
\usepackage{footnote}
\usepackage{color}

\let\oldfootsep=\footnotesep
\newcommand\ltsima{$\; \buildrel <\over\sim \;$}
\newcommand\simlt{\lower.5ex\hbox{\ltsima}}
\newcommand\gtsima{$\; \buildrel >\over\sim \;$}
\newcommand\simgt{\lower.5ex\hbox{\gtsima}}

\newcommand\msun {M_\odot}

\newcommand\mearth {{M_\oplus}}

 \def\t0{t_{\rm 0}}

\newcommand\pac{Paczy{\'n}ski }
\newcommand\ie{{\it i.e. }}

\newcommand\kpc {\, {\rm kpc}}

\shorttitle{}
\shortauthors{Bhattacharya et al}

\begin{document}

\title{Discovery of a Gas Giant Planet in Microlensing Event OGLE-2014-BLG-1760}


\author{A.~Bhattacharya\altaffilmark{1,M},
D.P.~Bennett\altaffilmark{1,2,M},
I.A.~Bond\altaffilmark{3,M},
T.~Sumi\altaffilmark{4,M},
A.~Udalski\altaffilmark{5,O},
R.~Street\altaffilmark{6,R}, 
Y.~Tsapras\altaffilmark{6,7,8,R}\\
  and\\
F.~Abe\altaffilmark{9}, M.~Freeman\altaffilmark{10}, A.~Fukui\altaffilmark{11}, Y.~Itow\altaffilmark{9}, M.~C.~A.~Li\altaffilmark{12}, C.~H.~Ling\altaffilmark{3}, K.~Masuda\altaffilmark{9}, Y.~Matsubara\altaffilmark{9},Y.~Muraki\altaffilmark{9}, K.~Ohnishi\altaffilmark{13}, L.~C.~Philpott\altaffilmark{14}, N.~Rattenbury\altaffilmark{12}, T.~Saito\altaffilmark{15}, A.~Sharan\altaffilmark{12}, D.~J.~Sullivan\altaffilmark{16}, D.~Suzuki\altaffilmark{2}, P.~J.~Tristram\altaffilmark{17}\\({\it MOA} collaboration),\\J. Skowron\altaffilmark{5}, M.~K.~Szyma\'nski\altaffilmark{5}, I.~Soszy\'nski\altaffilmark{5}, R.~Poleski\altaffilmark{5, 18}, P. Mr\'oz\altaffilmark{5}, S.~Kozlowski\altaffilmark{5}, P.~Pietrukowicz\altaffilmark{5}, K.~Ulaczyk\altaffilmark{5}, L.~Wyrzykowski\altaffilmark{5}\\({\it OGLE} collaboration),\\E.~Bachelet\altaffilmark{6,19}, D.~M.~Bramich\altaffilmark{19,20}, G. D'Ago\altaffilmark{21,22}, M.~Dominik\altaffilmark{23,\dagger}, R. Figuera Jaimes\altaffilmark{20,23}, K.~Horne\altaffilmark{23}, M.~Hundertmark\altaffilmark{24,25}, N.~Kains\altaffilmark{26}, J. Menzies\altaffilmark{27}, R. Schmidt\altaffilmark{7}, C.~Snodgrass\altaffilmark{28,29},I.~A.~Steele\altaffilmark{30}, J. Wambsganss\altaffilmark{7}\\({\it ROBONET} collaboration)
           } 
              
\keywords{gravitational lensing: micro, planetary systems}

\altaffiltext{1}{Department of Physics,
    University of Notre Dame, 225 Nieuwland Science Hall, Notre Dame, IN 46556, USA; 
    Email: {\tt abhatta2@nd.edu}}
\altaffiltext{2}{Code 667, NASA Goddard Space Flight Center, Greenbelt, MD 20771, USA}
\altaffiltext{3}{Institute of Natural and Mathematical Sciences, Massey University, Auckland 0745, New Zealand}
\altaffiltext{4}{Osaka University, 1-1 Yamadaoka, Suita, Osaka Prefecture 565-0871, Japan}
\altaffiltext{5}{Warsaw University Observatory, Al.~Ujazdowskie~4, 00-478~Warszawa, Poland}
\altaffiltext{6}{Las Cumbres Observatory Global Telescope Network, 6740 Cortona Drive, Suite 102, Goleta,CA 93117, USA}
\altaffiltext{7}{Astronomisches Rechen-Institut,  Zentrum f\"ur Astronomie der Universit\"at Heidelberg (ZAH), 69120 Heidelberg, Germany}
\altaffiltext{8}{School of Physics and Astronomy, Queen Mary University of London, Mile End Road, London E1 4NS, UK}
\altaffiltext{9}{Institute for Space-Earth Environmental Research, Nagoya University, 464-8601 Nagoya, Japan}
\altaffiltext{10} {Dept of Physics, University of Auckland, Private Bag 92019, Auckland, New Zealand}
\altaffiltext{11}{Okayama Astrophysical Observatory, National Astronomical Observatory of Japan, Asakuchi,719-0232 Okayama, Japan}
\altaffiltext{12}{Dept. of Physics, University of Auckland , Private Bag 92019, Auckland, New Zealand}
\altaffiltext{13}{Nagano National College of Technology, 381-8550 Nagano,Japan}
\altaffiltext{14}{Department of Earth, Ocean and Atmospheric Sciences, UNiversity of British Columbia, Vancouver, British Columbia, Vancouver, British Columbia, V6T 1Z4, Canada}
\altaffiltext{15}{Tokyo Metroplitan College of Industrial Technology, 116-8523 Tokyo, Japan}
\altaffiltext{16}{School of Chemical and Physical Sciences, Victoria University, Wellington, New Zealand}
\altaffiltext{17}{Mt. John University Observatory, P.O. Box 56, Lake Tekapo 8770, New Zealand}
\altaffiltext{18}{Department of Astronomy, Ohio State University, 140 West 18th Avenue, Columbus, OH 43210, USA}
\altaffiltext{19}{Qatar Environment and Energy Research Institute (QEERI), HBKU, Qatar Foundation, Doha, Qatar}
\altaffiltext{20}{European Southern Observatory, Karl-Schwarzschild-Str. 2, 85748 Garching bei M\"unchen, Germany}
\altaffiltext{21}{Dipartimento di Fisica ``E.R. Caianiello", Universit\`a di Salerno, Via Ponte Don Melillo, 84084-Fisciano (SA), Italy}
\altaffiltext{22}{Istituto Nazionale di Fisica Nucleare, Sezione di Napoli, Napoli, Italy}
\altaffiltext{23}{School of Physics \& Astronomy, University of St Andrews, North Haugh, St Andrews KY 16 9SS, UK}
\altaffiltext{24}{Niels Bohr Institute, University of Copenhagen, Juliane Maries Vej 30, 2100, Kobenhavn, Denmark}
\altaffiltext{25}{SUPA, School of Physics \& Astronomy, University of St Andrews, North Haugh, St Andrews KY16 9SS, UK}
\altaffiltext{26}{Space Telescope Institute, 3700 San Martin Drive, Baltimore, MD 21218, USA}
\altaffiltext{27}{South African Astronomical Observatory, PO Box 9, Observatory 7935, South Africa}
\altaffiltext{28}{Planetary and Space Sciences, Dept of Physical Sciences, The Open University, Milton Keynes, MK7 6AA, UK}
\altaffiltext{29}{Max Planck Institute for Solar System Research, Justus-von-Liebig-Weg 3, 37077 Gotingen, Germany}
\altaffiltext{30}{Astrophysics Research Institute Liverpool John Moores University, Liverpool L3 5RF, UK and Royal Society University Research Fellow}
\altaffiltext{$\dagger$}{Royal Society University Research Fellow}
\altaffiltext{M} {Microlensing Observation in Astrophysics}
\altaffiltext{O} {Optical Gravitational Lensing Experiment}
\altaffiltext{R} {Robonet}

\newpage


\begin{abstract}
We present the analysis of the planetary microlensing event OGLE-2014-BLG-1760, which shows a strong 
light curve signal due to the presence of a Jupiter mass-ratio planet. One unusual feature of this event is
that the source star is quite blue, with $V-I = 1.48\pm 0.08$. This is marginally consistent with source star
in the Galactic bulge, but it could possibly indicate a young source star in the far side of the disk. 
Assuming a bulge source, we perform a Bayesian analysis assuming a standard Galactic model, and
this indicates that the planetary system resides in or near the Galactic bulge at 
$D_L = 6.9 \pm 1.1{\rm \kpc}$. It also indicates a host star mass of $M_* = 0.51 \pm 0.44\msun$,
a planet mass of $m_p = 180 \pm 110\mearth$, and a projected star-planet separation
of $a_\perp = 1.7\pm 0.3\,$AU. The lens-source relative proper motion is 
$\mu_{\rm rel} = 6.5\pm  1.1$ mas/yr. The lens (and stellar host star) is predicted to be very faint, so 
it is most likely that it can detected only when the lens and source stars are partially resolved. Due to the relatively
high relative proper motion, the lens and source will be resolved to about $\sim46\,$mas in 6-8 years 
after the peak magnification. So, by 2020 - 2022, we can hope to detect the lens star with deep, 
high resolution images.       
\end{abstract}

\section{Introduction}
\label{sec-intro}
Gravitational Microlensing is the technique of detecting exoplanets using gravitational lensing. 
Microlensing is unique in its ability to detect planets \citep{gouldloeb1992} just outside of the snow line \citep{lissauer_araa} down to 
an earth mass \citep{bennett1}, which is difficult or impossible with other methods. According to the 
core accretion theory, the snow line \citep{snowline1,snowline2, snowline3, kenyon_hartmann} plays a very 
crucial role in the planet formation process. Beyond the snow line, ices condense, increasing the 
density of the solid materials. Higher density of solids speeds the planet formation process in the 
protoplanetary disk, hence forming cold planets quickly beyond the snow line. The planets, 
discovered by microlensing, provide the statistics needed to understand the architecture of cold planets 
beyond the snow line \citep{suzuki2016}. Since this method does not rely on the light from 
the host stars, it can detect planets, even when the host stars cannot be detected \citep{maopaczinski}. 
Most of the planets discovered so far using microlensing are $\sim 1$-$8\,$kpc away. 
Thus, microlensing is able to detect planets in the inner Galactic disk and bulge, where it is difficult
to detect planets with other methods. Thus, microlensing has the potential to measure how the 
properties of exoplanets depend on the Galactic environment.

For most planetary microlensing events, the angular Einstein radius, $\theta_{E}$ is measured 
from the finite source effect. In events where the lens star brightness \citep{bennett06} or the 
parallax effect is measured \citep{parallax1,gaudi2008,muraki}, the planetary mass and distance to 
the planetary system can be determined . Hence this technique can be used to build statistics of
planetary mass as a function of the host star mass. Since the planets detected
by microlensing are $\sim 1$-$8\,$kpc away, the distance to the planetary system will also allow
the determination of the planetary mass function as a function of the distance towards the galactic center.      

In this paper we present the discovery of a gas giant planet orbiting the lens stars for microlensing
event OGLE-2014-BLG-1760. The mass ratio of this planet is $q = 8.64 \times 10^{-4}$, which is 
slightly less than that of Jupiter. The paper is organized as follows: Section \ref{sec-lc_data} describes 
the light curve data collected for the event OGLE-2014-BLG-1760. The next section (Section \ref{sec-lc}) 
is divided in four parts: \ref{sec-Data-reduction} summarizes data reduction procedures for 
the different data sets; Section \ref{sec-Best-fit} shows the best fit model and procedures that are 
used to obtain it; Section \ref{sec-Limb-darkening} describes how the limb darkening of the source 
is modeled; and section \ref{sec-parallax} presents our attempt to detect the microlensing parallax
effect in the light curve. In Section \ref{sec-radius}, we discuss source brightness and angular
radius measurement and derive the lens-source relative proper motion. 
Section \ref{sec-Lens} discusses an estimate of the lens properties and the future possible investigations. 

\section{Observation}
\label{sec-lc_data}

The {\it OGLE} (Optical Gravitational Lensing Experiment) collaboration operates a 
microlensing survey towards the galactic bulge, with the 1.3 meter Warsaw telescope from 
Las Campanas observatory in Chile. 
Most of the OGLE-IV (phase 4) observations were taken in the Cousins $I$-band, with occasional
observations in Johnson $V$-band \citep{ogledata}. In the year 2014, OGLE Early Warning 
System (EWS) \citep{ogle-ews} has alerted 2049 microlensing candidates of which this event 
OGLE-2014-BLG-1760 was the 1760th one. 

The  {\it MOA} (Microlensing Observation in Astrophysics) \citep{moapipeline} collaboration also 
operates a microlensing survey towards the galactic bulge with the 1.8 meter MOA-II telescope 
from Mount John Observatory at Lake Tekapo, New Zealand. 
The observations are mostly taken in  MOA-red wide band filter which covers the wavelengths of the standard 
Cousins $R + I$ bands. For year 2014, MOA has reported about $\sim$33 microlensing anomalies 
out of which OGLE-2014-BLG-1760 planetary event is one of them.

Both of these telescopes have relatively large fields-of-view, 2.2 deg$^{2}$ for MOA-II, and 
1.4 deg$^{2}$ for OGLE-IV. These enable survey observations with cadences as high as
one observation every 15 minutes, and this allows the surveys to detect the sharp light curve
features of planetary light curve features, when they are only smoothed by the finite source effects
of a main sequence source star.
It is the high cadence observation of microlensing events of MOA-II and OGLE-IV survey that helps in detecting microlensing anomalies, including the microlensing planetary signatures.

The microlensing event OGLE-2014-BLG-1760 was discovered at (RA,decl.)(2000) 
= ($17:57:38.16,-28:57:47.37$)[(l,b)=($1.3186,-2.2746$)], by the OGLE EWS on August 22, 2014 
around 7.25 a.m. EDT. The same event was alerted by MOA on August 31, 2014 as MOA-2014-BLG-547. 
Later, on September 10, 2014, the MOA collaboration detected the planetary cusp crossing and 
announced the anomaly in the light curve (around HJD = 2456911 in Figure 1). In response to 
MOA anomaly alert, the follow up groups Robonet and $\mu$FUN started collecting data on this event.
The high cadence observation of MOA-II survey covered most of the cusp crossing, and the trough after 
the cusp crossing was well covered by follow up groups, Robonet and $\mu$FUN. 
The Robonet group observed the event with 1 m robotic telescopes at Cape Town, South Africa 
and at Siding Springs, Australia, in the Sloan $I$-band. The $\mu$FUN group also observed the event 
with the 1.3 m SMARTS CTIO telescopes in both the $V$ and $I$-bands. All these observations are 
shown in Figure 1.  Although the $\mu$FUN data is pretty flat because the trough
is followed by the normal single lens decrease ( see Figure 1), the $\mu$FUN data do help
to exclude the wide model with s = 1.27 and contribute in constraining the planetary model ( discussed in section \ref{sec-Best-fit}). It is true that the constraint
from the Robonet and OGLE data is stronger because of better Robonet
coverage and OGLE observations over a wider range of magnifications.
\begin{figure}
\epsscale{1.0}
\plotone{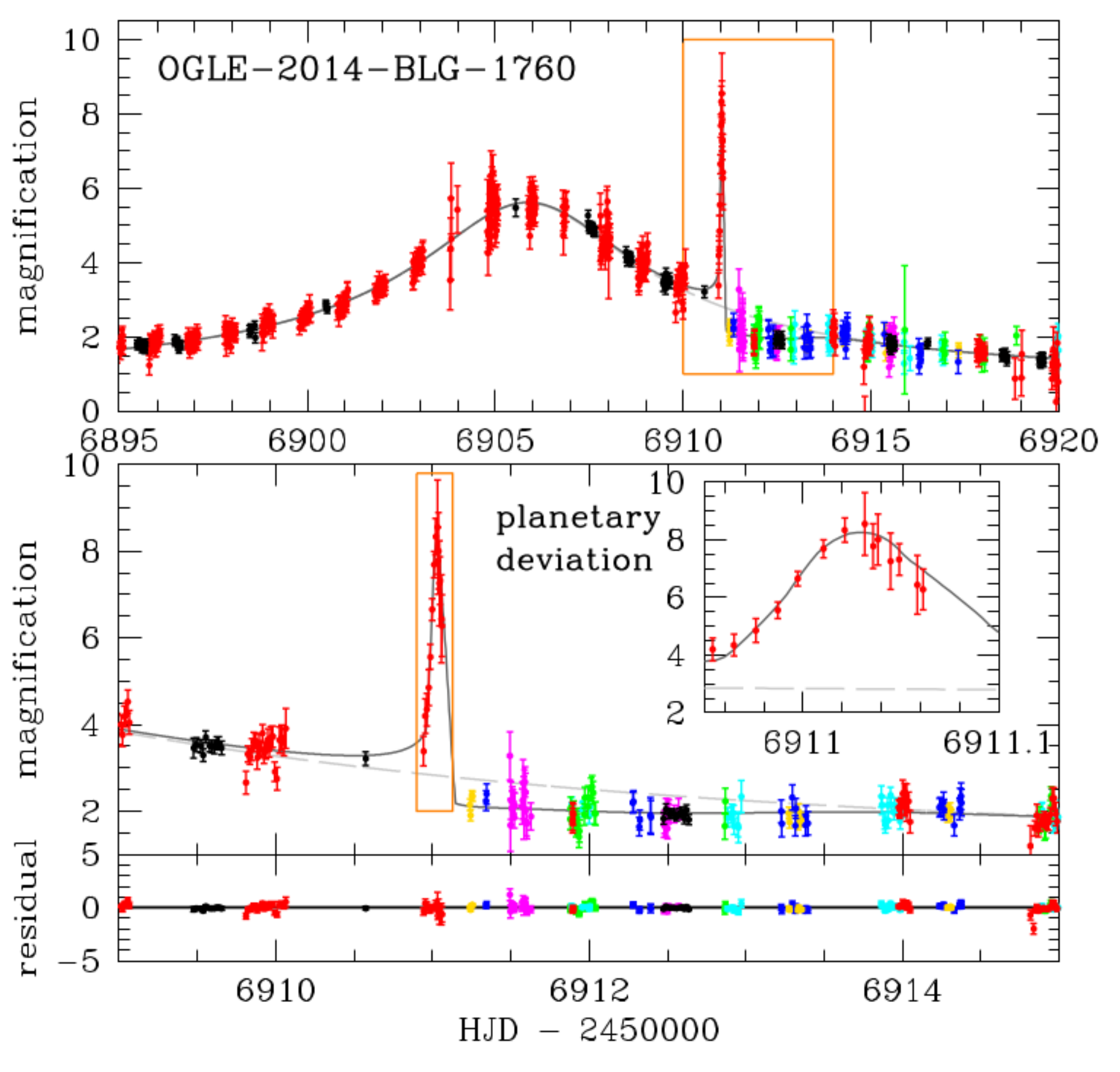}
\caption{The light curve of event OGLE-2014-BLG-1760. The light grey dashed line represent the 
best fit single lens model and the dark grey line represents the best planetary model. The data are
plotted in the following colors: red for the MOA-red band, black for OGLE-$I$, purple for OGLE-$V$,
(difficult to see due to overlap with the OGLE-$I$ data), blue for the ROBONET SAAO-A $I$-band, gold 
for the ROBONET SAAO-B $I$-band, green for the ROBONET Siding Springs-A $I$-band, 
cyan for the ROBONET Siding Springs-B $I$-band, and magenta for the $\mu$FUN $I$-band.  
\label{fig-lc}}
\end{figure}
\begin{figure}
\label{fig-caus}
\epsscale{1.0}
\plotone{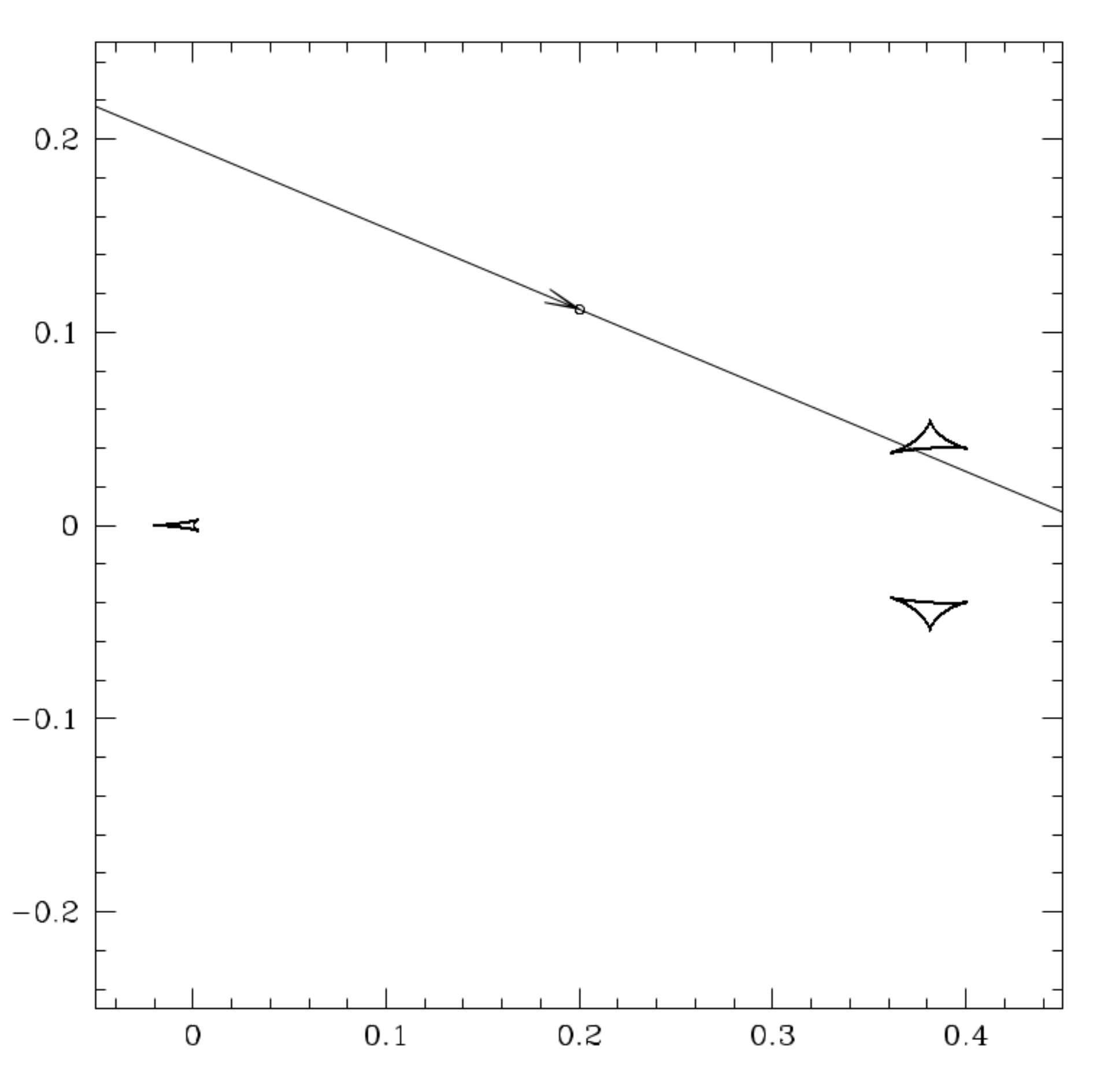}
\caption{This figure shows the caustic configuration and source trajectory for the best fit planetary model. 
When the source crosses inside the planetary caustic, its magnification jumps as two new highly magnified 
images are created, and then the magnification drops just as abruptly when the source exits and
two images disappear. The two caustic crossings are at HJD$^\prime - 2450000 = 6911.00$ and
$6911.07$. Since the interval between the caustic crossings is similar to the source radius crossing
time, $t_*$, the two magnification peaks merge into one peak, as shown in Figure \ref{fig-lc} at 
HJD$^\prime \sim 6911$. Also, when the source passes between the two minor image
planetary caustics, the magnification of the source brightness drops, as can be seen in the light curve 
after caustic crossing, between HJD$^\prime$ = 6911.20 and 6914.00 in Figure \ref{fig-lc}. }
\end{figure}

\section{Data Reduction and Modeling}
\label{sec-lc}
\subsection{Data Reduction}
\label{sec-Data-reduction}
All the images were reduced to photometry using the difference image method \citep{dia-3,dia-2,dia-1}. The MOA images were reduced using MOA difference image analysis pipeline (\cite{ moapipeline}). 
The data was then detrended using the non-2014 data to minimize the error due to the effects of 
differential refraction, seeing and airmass. To avoid systematic errors due to flat field changes and 
changes in detector sensitivity, we removed data prior to 2011 and only use the data after 
HJD$^\prime = 5596$ (February, 2011) for modeling. OGLE data were reduced using OGLE Difference 
Image Analysis software and optimal centroid method (\cite{oglepipeline,oglepipeline1}). 
The $\mu$FUN photometry was produced using a modified version of the PySIS package \citep{pysis}. 
Robonet data were reduced using the DanDIA package \cite{dandia}. 

The error bar estimates obtained from the photometry codes are sufficient to find the best fit 
model. These error bar estimates are good measures of the relative uncertainties for measurements 
with the same telescope, but they are often wrong by a factor of $\sim$2. To determine the 
uncertainties for the model parameters, it is necessary to have accurate error bars which will give 
$\chi^2/{\rm dof}\approx 1$ for each passband. We used method of \citet{bennett14} to normalize 
the error bars using $\sigma^\prime_{i} = k \sqrt{\sigma^{2}_{i}+e_{min}^2}$. Here $\sigma_{i}$ is the 
initial uncertainty of i-th data point; $\sigma^\prime_{i}$ is the modified error bar. A $k$ value is selected
for each passband to give $\chi^2/{\rm dof} = 1$. Our initial fits used $k = 1.50$ and $e_{min} = 0.003$ 
for all passbands. Next, once the best fit model was found, the values of $k$ was modified for each 
passband to satisfy $\chi^2/{\rm dof} = 1$. The number of datapoints used for each passband and 
their corresponding $k$ values are listed in Table \ref{Data-red}. New fits were done with the new error
bars, which resulted in very small changes to the model parameters. 
\begin{deluxetable}{cccc}
\tablecaption{Data Reduction Summary\label{Data-red}}
\tablewidth{0pt}
\tablehead{\colhead{Telescope}&\colhead{Filter}&\colhead{N$^a$/N$^b$}&\colhead{$k$}}
\startdata
MOA &red(I+R)&11655$/$11673& 0.63\\
OGLE&I&10779$/$10794&0.98\\
OGLE&V&119$/$119&0.84\\
$\mu$FUN&I&45$/$45&0.79\\
$\mu$FUN&V&3$/$4& 0.99\\
ROBONET SAAO A&I&54$/$54&1.16\\
ROBONET SAAO B&I&32$/$32&0.96\\
ROBONET Siding springs- A&I&56$/$56&1.24\\
ROBONET Siding springs- B&I&75$/$75&1.23\\
\enddata
\\$^a$ Number of data points used\\$^b$ Number of total observations
\end{deluxetable}

\subsection{Best Fit Model}
 \label{sec-Best-fit}
We begin our modeling of the OGLE-2014-BLG-1760 light curve with single lens microlensing model
\citep{pac1986}. There are three non-linear model parameters for a single lens event: 
$t_0$ - the time of peak magnification, $u_0$ - the minimum separation between source and lens 
in Einstein radius units, and $t_E$ - the Einstein radius crossing time. There are also two linear 
parameters for each passband: the source flux $f_{s}$,  and the blend flux, $f_{bl}$. We find the 
best single lens parameters as the starting point for a systematic search through parameter space
to find the best binary lens solution. Because the light curve follows a single lens shape for most
of its history, except in the vicinity of the planetary feature at $t \approx 6911$, we can use the best fit
single lens parameters as the starting point for an initial condition grid search, following \citet{bennettmcmc}.

To describe a binary lens, we need three additional parameters: the lens mass ratio, $q$, the 
projected separation between the lens masses, $s$, measured in Einstein radius units, and
the angle between the source trajectory and the lens axis, $\theta$. Also, binary events often
have caustic or cusp crossings, which resolve the angular size of the source, so we need an additional
parameter, the source radius crossing time, $t_*$, to model finite source effects. With the single
lens parameters fixed to the best fit values, we using the initial condition grid search method to 
search over the parameter ranges $0.48 \leq s \leq 2.10$, $-4 \leq \log q \leq -2 $, and 
$ -\pi \leq \theta \leq \pi$, with $t_*$ fixed at $t_* = 0.05$. We then select $\sim 10$ of the best
fit values from the initial condition grid search (with very different values of $q$, $s$ and $\theta$) to
use as initial conditions for full, non-linear modeling runs using the  \citet{bennettmcmc} $\chi^{2}$ 
minimization recipe, which is a modification of Markov Chain Monte Carlo algorithm \citep{mcmc}.
The parameters of the best fit binary lens model and the $\chi^2$ improvement are compared 
with the best fit single lens model in Table \ref{tab-params}. The binary lens best fit model has 
$q = 8.64 \times 10^{-4}$ improves the renormalized $\chi^2$ by $\Delta\chi^2 = 1218.75$ 
compared to the best fit single lens model. Hence the binary lens model is the preferred model. 
 
High magnification planetary microlensing events usually have a ``close-wide" degeneracy, in that 
solutions with $s \leftrightarrow 1/s$ are nearly degenerate. This is usually not the case for low
magnification planetary events because the caustic structure for the major and minor image 
caustics is quite different. However, the major and minor image caustics with $s \leftrightarrow 1/s$
are encountered at the same single lens magnification, and with specific source trajectories, it
is possible to produce similar light curves with both $s > 1$ and $s < 1$, particularly if the light
curves aren't very well sampled. Therefore, we have searched for models with $s \sim 1/0.83 =1.20$.
We found a best fit $s > 1$ model with $s = 1.27$, but this is a worse fit than the best fit model
(with $s = 0.83$) by $\Delta\chi^2\sim 215$. This is because the OGLE-2014-BLG-1760 is 
well sampled, so the $s = 1.27$ model is excluded, and the $s=0.83$ is the only viable
solution.

\subsection{Limb Darkening Effect}
\label{sec-Limb-darkening}
The photometric calibrations and the extinction toward the red clump stars in the vicinity of the 
source are discussed in Section~\ref{sec-radius}. The magnitude and color of the source 
are indicated with the cyan point in Figure \ref{fig-cmd}, and the extinction corrected color is 
$(V-I)_{\rm S,0} = 0.34$, as discussed in section \ref{sec-radius}. This color implies that
the source is an $\sim$A9 star with $T_{\rm eff} \sim 7352$ K from \citet{kenyon_hartmann}.
We use a linear limb darkening model, and from \citet{claret2000}, we select the limb darkening parameters 
$u$ to be 0.4204, 0.5790 and 0.46035 corresponding to the $V$, $I$ and MOA-Red bands, respectively.
These corresponds to the temperature of $T_{\rm eff}= 7352$ K and a surface gravity of $\log g = 4.5$. The parameters of the best fit model are insensitive to the precise limit darkening parameters.

\subsection{Search for a Microlensing Parallax Signal}
\label{sec-parallax}
The microlensing parallax effect has been detected in a number of planetary microlensing events
\citep{gaudi2008,bennett2010,muraki} where it has allowed the lens system masses 
to be measured. Due to the Earth's orbital motion about the Sun, the apparent lens - source relative 
motion deviates from uniform linear motion. This phenomenon is known as microlensing parallax 
effect \citep{parallax1,parallax2}. The parallax effect can be described with the parallax vector 
$\boldsymbol{\pi_E} = (\pi_{E,N},\pi_{E,E})$, where direction of $\boldsymbol{\pi_E}$ is same 
as lens-source relative proper motion, $\mu_{\rm rel}$. The amplitude of $\pi_E$ is the inverse 
of the Einstein radius projected to the observer's plane, $\pi_E = {\rm AU} / \tilde{r}_E$. 
When both the microlensing parallax effect and the angular Einstein radius (as described in 
section \ref{sec-radius}) are measured, we can determine the lens system mass and distance from
the following equations: 
\begin{eqnarray}
M_{\rm L} = \frac{\theta_E}{k\pi_E} \ \ \ \ \ \ \ \ \ \ \   \\
D_{\rm L} = \frac{\rm AU}{\pi_E\theta_E + \frac{\rm AU}{D_{\rm S}}}  \ ,
\end{eqnarray}   
where $k = 4G/(c^2 {\rm AU}) = 8.14\,{\rm mas}\, \msun^{-1}$,
$M_{\rm L}$ is the total mass of host star and planet and $D_{\rm L}$ and  $D_{\rm S} \sim 8\,$kpc
are the distances to the lens system and source stars, respectively. The host star mass, $M_*$,
is given by $M_* = M_{\rm L}/(1+q)$.

\begin{figure}
\epsscale{1.0}
\plotone{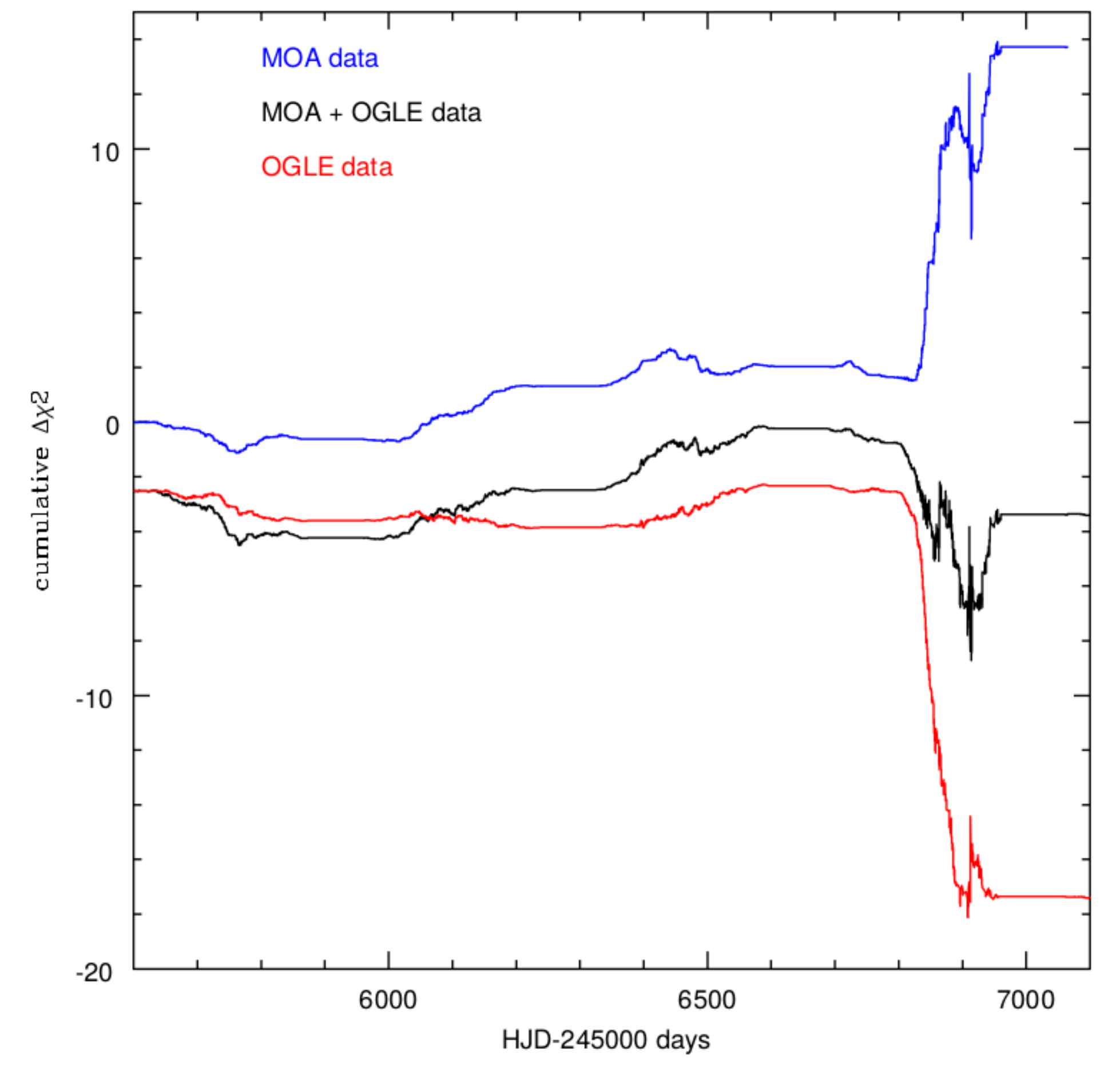}
\caption{The difference in the cumulative $\chi^2$ values between the best fit models with
and without parallax are shown for MOA data (blue), OGLE data (red) and total MOA + OGLE data (black). 
The negative cumulative $\Delta\chi^2$ means parallax model is supported. Also MOA data does not support parallax whereas OGLE data supports parallax model. The $\chi^2$ improvement from MOA + OGLE data for parallax model is $\sim 3.60$. The overall $\chi^2$ improvement for parallax model from all the telescopes is $\sim 9.90$. Hence the parallax signal is dubious and probably a false signal.  
\label{fig-par}}
\end{figure}


In order to include the microlensing parallax effect, we must add two new model parameters 
$\pi_E$ and $\phi_E$, the magnitude and direction angle of the parallax vector. (The north and east
components of $\boldsymbol{\pi_E}$ are given by $\pi_{E,N} = \pi_E \cos(\phi_E)$ and 
$\pi_{E,E} = \pi_E \sin(\phi_E)$.) Our best fit parallax model has an unusually large $\pi_E$
value of $\pi_E=5.86$, which would imply a very nearby and low mass lens if it is an accurate measurement,
but the improvement in $\chi^2$ is only $\Delta\chi^{2}=9.90$ over the best fit without parallax.
We can examine the origins of this parallax signal by examining the cumulative $\Delta\chi^2$
between the non-parallax and parallax models as a function of time, which we show in 
Figure~\ref{fig-par}. The cumulative $\Delta\chi^2$ is displayed for the MOA and OGLE data
separately, and we can see that only the OGLE data favors the parallax model.
\citet{penny2016} consider the distribution of published planetary microlensing events and
show the there an implausibly large number with ``high" microlensing parallax values 
($\pi_E \simgt 1$). This suggests that some of the published planetary events have spurious
large $\pi_E$ values \citep{han_ob130723}. We conclude that this microlensing parallax
measurement for OGLE-2014-BLG-1760 is also spurious.
 
 
\begin{deluxetable}{cccc}
\tablecaption{Model Parameters \label{tab-params}}
\tablewidth{0pt}
\tablehead{\colhead{parameter}&\colhead{units}&\colhead{binary lens best fit}&\colhead{single lens best fit}}
\startdata
$t_E$& days &15.87 $\pm$ 0.41&15.47\\
$t_0$&HJD$-2450000$&6905.856 $\pm$0.026&6905.9541\\
$u_0$&&0.1806 $\pm$ 0.0074&0.19\\
$s$ &&0.8269 $\pm$ 0.0047&\\
$\theta$&radian&-0.3977 $\pm$0.0086&\\
$q$&$10^{-4}$&8.64 $\pm$ 0.89& \\
$t_\star$&days&0.0366 $\pm$0.0044&\\
fit $\chi^{2}$&&22818.05 &24036.80\\
\enddata
\end{deluxetable}   

\section{Calibration and Source Properties}
\label{sec-radius}
The OGLE data were taken in the OGLE-IV $I$ and $V$-bands, which we calibrate to 
the OGLE-III catalog Cousins $I$ and Johnson $V$ band \citep{ogleiii}. 151 bright 
($I\leq 16.50$) and isolated stars were matched in both passbands and used
for this calibration. The following calibration relations are used to convert OGLE $VI$ 
magnitudes to OGLE III Catalog Cousins $I$ and Johnson $V$ magnitude:

\begin{eqnarray}
I_{\rm OGLE_{IV}} = -0.0471 + 0.99867 I_{\rm OGLE_{III}} + 0.00133 V_{\rm OGLE_{III}} \\
V_{\rm OGLE_{IV}} = -0.3444 - 0.10068 I_{\rm OGLE_{III}} + 1.10068 V_{\rm OGLE_{III}}
\end{eqnarray}   
The uncertainty in the brightness due to these calibration relations is $\sim 0.01$ magnitude. 
The source brightness in the OGLE III catalog scale is $I_{\rm S} = 19.07 \pm 0.14$ , 
$V_{\rm S} = 20.51 \pm 0.26$. The errors in brightness are calculated from MCMC averages 
over all the MCMC fits and from the uncertainty in the calibration relations. 

\begin{figure}
\epsscale{1.0}
\plotone{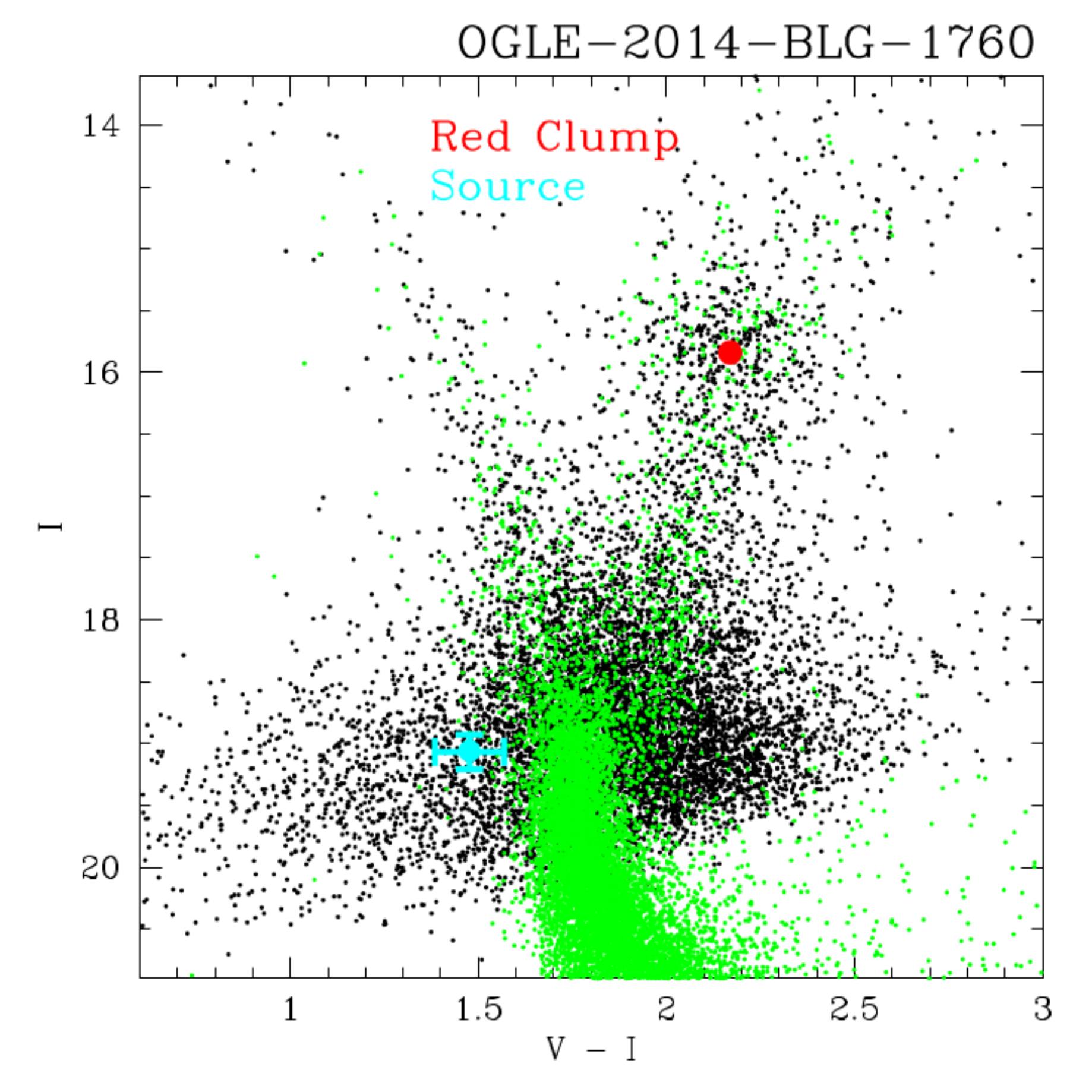}
\caption{The $(V-I, I)$ color magnitude diagram (CMD) of the stars in the OGLE-III catalog
\cite{ogle3phot}
within $140^{\prime\prime}$ of OGLE-2014-BLG-1760. The red spot indicates
red clump centroid, and the cyan spot indicates the source magnitude
and color with error bars. 
The green points show HST CMD of Baade's window transformed to have the same red clump
centroid as this field. The source star is slightly bluer than the HST main sequence stars.
This might be explained by a young, blue source in the Milky Way disk on the far
side of the bulge.} 
\label{fig-cmd}
\end{figure}
   
There is a single magnified OGLE $V$-band observation at $t = 6896.50$ (HJD$^\prime$),
and this results in a relatively large, $10\%$ uncertainty in the $V$-band source flux
for the best fit model. As a result, the OGLE $VI$ source colors given a source star color
of $V-I = 1.45 \pm 0.11$. The color of the source star is bluer than the average main 
sequence star color as can be seen in figure \ref{fig-cmd}. Because we have only this single
$V$-band measurement, we also calculate the source star color from MOA-Red and OGLE-$I$ 
band following \citet{gould_col} and \citet{bennett12}. Since we have a large number of 
MOA-Red and OGLE-$I$ observations when the source is significantly magnified, we expect
a robust measurement of the $ R_{\rm MOA} -I$ source color. About 140 bright, isolated 
stars with $I$-band magnitude $I <$ 16 and $1.0 < V-I < 2.6$ are matched between the
MOA-II and OGLE III reference images, and we used these to find the following calibration relation:
  $$ R_{\rm MOA} -I_{\rm OGLE_{III}} =  0.18143\times (V-I)_{\rm OGLE_{III}} \ , $$ 
where the $R_{\rm MOA}$ passband has been calibrated to give $R_{\rm MOA} = I_{\rm OGLE_{III}}$
when $(V-I)_{\rm OGLE_{III}} = 0$. The RMS error for this calibration relation is 0.0343,
for a formal uncertainty of $0.0343/\sqrt{N} = 0.0029$, but we assume a calibration uncertainty
of 0.02 for $ R_{\rm MOA} -I_{\rm OGLE_{III}}$.

Our models give a source brightnesses of $I_S = 19.07 \pm 0.14$ and 
$R_{{\rm MOA},S} = 19.34 \pm 0.15$ and a calibrated source color of $V-I =1.52\pm 0.11$.
This color is less than 1-$\sigma$ away from the color of the source derived from OGLE $V$-band 
data, so we conclude that the blue color from the OGLE data is probably real. However, the average
color from both measurements is $V-I = 1.48 \pm 0.08$, which still has a relatively large error bar.
We take this average as the correct color of the source. This color is plotted in blue point in the
color magnitude diagram of all stars within $140^{\prime\prime}$ of the source star, 
Figure \ref{fig-cmd}. The green points in the CMD figure \ref{fig-cmd} 
show the HST CMD plot \citep{hstcmd} shifted to have the same red clump centroid position 
as the OGLE-III stars . The source color is slightly bluer than the main sequence star 
distribution shown with green points in the HST CMD in figure \ref{fig-cmd}. The average color of main 
sequence bulge stars in the same magnitude range of the source ($I = 19.07 \pm 0.14$) in
in the HST CMD is $V-I = 1.75$. We can estimate the probability that a star with the measured color 
of the source ($V-I = 1.48 \pm 0.08$) is drawn from this HST distribution by integrating the 
Gaussian describing the measured source color with its error bar over the distribution of stars
in the same magnitude range from the HST CMD. We then divide this result by the result of
the integral with the same error bar, but centered on the average color of $V-I = 1.75$. The ratio of
integral centered on the measured color and the integral centered on the average color is 0.05,
indicating that this source color is marginally consistent with being drawn from the known
bulge star population. With nearly, 50 planetary microlensing events published to date, we would
expect one or two to be found with such a blue source star. On the other hand, it is also possible
that blue stars are preferentially lensed because they are at a greater distance. This can be the
case if the source is a young blue (late A or early F) star on the far side of the Galactic disk,
assuming negligible extinction beyond the bulge on this line-of-sight.
Such stars would have a much larger microlensing event rate, so they could be preferentially
lensed by bulge stars. On the other hand, it is possible that the CMD in this field is noticeably 
different than that of Baade's window where the \citet{hstcmd} image was taken.
 
We calculate the extinction in the direction of the source using the method described by 
\citet{bennett14}. The position of the centroid of red clump in the OGLE III catalog is found to
be $(V-I, I)_{\rm RC}$ = (2.20, 15.84) as shown in figure \ref{fig-cmd}. From \citet{bensby} and 
\citet{nataf}, the dereddened red clump centroid is determined to be 
$(V-I, I)_{\rm RC}$ = (1.06, 14.39). Hence, the extinction at this galactic coordinate is :
$(E(V-I), A_I)_{\rm RC}$ = (1.14, 1.45). The extinction to the source star is assumed to be 
same as the average extinction for the red clump stars within $140^{\prime\prime}$ since 
most of the extinction is thought to be in the foreground. Thus, the extinction corrected source 
color and brightness are $(V-I, I)_{\rm S, 0}=$ (0.34, 17.62), as shown in blue in figure \ref{fig-cmd}.

The source radius is calculated from the dereddened source magnitude and color using the following 
formula, obtained from a private communication with Tabetha Boyajian:
\begin{equation}
\log_{10}(2\theta_*) = 0.50141358 + 0.41968496 (V - I) -0.2 I
\end{equation}
For the source color or $V-I = 1.48$, the calculated source radius is 
$\theta_* =$ (6.57 $\pm$ 0.11) $\times$ 10$^{-4}$ mas. The relative proper motion and Einstein 
radius are calculated from:
\begin{eqnarray}
\mu_{\rm rel} = \frac{\theta_*}{t_{\star}}\\
\theta_{E} = \theta_* \times \frac {t_{E}}{t_{\star}}
\end{eqnarray}     
Since the best fit model has $t_E = 15.87$ days and $t_* = 0.04$ days, we find:
\begin{eqnarray}
 \mu_{\rm rel} = 6.55 \pm 1.12 {\rm ~mas/yr}\\
 \theta_E = 0.29 \pm 0.05 {\rm ~mas} \ .
 \end{eqnarray}    
There is a $5\%$ uncertainty in source radius relation (equation 5). We assume 1-2$\%$ 
error in the calibration relations (equations 3 and 4).
 
\section{Lens Properties and Discussion}
\label{sec-Lens}

  \begin{deluxetable}{ccc}
\tablecaption{Lens System Parameters \label{tab-params-histogram}}
\tablewidth{0pt}
\tablehead{\colhead{parameter}&\colhead{units}&\colhead{$(V-I)_S = 1.48\pm 0.08$}}
\startdata
Host star mass, $M_*$&${\msun}$&0.51$^{+0.44}_{-0.28}$\\
Planet mass, $m_{\rm P}$&$\mearth$&182$^{+137}_{-83}$\\
Host star - Planet 2D separation, $a_{\perp}$&AU&1.75$^{+0.34}_{-0.33}$\\
Host star - Planet 3D separation, $a_{3D}$&AU&2.57$^{+1.12}_{-0.45}$\\
Lens distance, $D_{\rm L}$&\kpc &$6.86\pm 1.11$\\
Lens magnitude, $I_{Lens}$&Cousins $I$&23.42$^{+1.89}_{-2.92}$\\
Lens magnitude, $H_{Lens}$&$H$&20.80$^{+1.54}_{-2.31}$\\
\enddata
\end{deluxetable}
If we assume a standard Galactic model \citep{bennett14} with a source star in the bulge
and we assume that all stars and brown dwarf have an equal probability of hosting a 
planet with the measured properties, then we can perform a Bayesian analysis to estimate 
the lens system properties. We ran an MCMC run with about 180,000 links to obtain the 
{\it posterior} distributions presented in Table \ref{tab-params-histogram} and Figure 5.
The mass of the host star is only approximately determined to be $M_* = 0.51{+0.44\atop -0.28} \msun$,
so it could be an M, K, or G star. The 1-sigma range of the planet mass, 
$m_p = 182{+137\atop -83}$ spans the range from the mass of Saturn to that of Jupiter.
The distance to the lens system is more precisely determined, due to the relatively small
angular Einstein radius, $\theta_E = 0.29 \pm 0.05\,$mas. Our analysis predicts a
lens system distance of $D_L = 6.86 \pm 1.11\,$. This implies that the lens system
is very likely to be in the Galactic bulge.

\citet{penny2016} argue that the published planetary microlensing events show a dearth of 
planets orbiting Galactic bulge stars. This OGLE-2014-BLG-1760L lens system would seem 
to be  a counterexample, along with a number of other planetary microlens systems,
such as OGLE-2005-BLG-380L \citep{ogle390}, OGLE-2008-BLG-092L \citep{poleski_ob08092},
MOA-2008-BLG-310L \citep{janczak}, OGLE-2008-BLG-355L \citep{koshimoto14},
MOA-2011-BLG-353L \citep{rattenbury2015} and MOA-2011-BLG-293L \citep{yee2012,batista14}.
Actually, the problem with the \citet{penny2016} analysis is pretty easy to understand. The
Einstein radius crossing times ($t_E$) for bulge events are significantly smaller than the $t_E$ values
for disk events, but the detection efficiencies for microlensing events and planetary signals
are substantially higher for events with large $t_E$ values. The statistical analysis of
\citet{suzuki2016} shows that the planet detection efficiency weighted median $t_E$ 
value is $\approx 42\,$ days. This effect
occurs for high magnification events \citep{gould10} because high magnification is much
easier to predict when $t_E$ is large, and for low magnification events \citep{suzuki2016} 
because the planetary signal duration (for fixed $q$) is proportional to $t_E$. This effect
can be accounted for with accurate detection efficiency calculations, but \citet{penny2016} 
use a detection efficiency calculation for an advance ground-based survey, that is
much more sensitive than any ground-based survey undertaken to date. One of us
\citep{bennett04} has previously studied a variety of ground-based microlensing surveys
with a wide range of telescope options, and these simulations show that this detection
efficiency bias with $t_E$ is much stronger with the less capable surveys that can 
approximate the sensitivity of the observing programs that have discovered the 
published events.
  
It should also be noted that the color of the source star $V-I = 1.48\pm 0.08$ implies 
that this star is bluer than and only marginally consistent with bulge main sequence
shown in Figure~\ref{fig-cmd}. This source star could be a "blue straggler" which is 
an old, bright blue star that probably formed from a merger of two smaller stars, but 
these stars are rare, or it could be that this field has a larger fraction of bluer, metal
poor stars than the HST Baade's window stars plotted in Figure~\ref{fig-cmd}. But this
also seems unlikely. Another possibility is that this star is a relatively young blue star that
resides in the Galactic disk on the far side of the bulge with minimal additional extinction
between the bulge and this source star. If so, HST observations taken separated
by 2 years could reveal the characteristic source star proper motion of a far side disk star. 
Since the galactic coordinates of the source, $(l,b) = 1.3186, -2.2746$ are pretty 
close to the nominal WFIRST exoplanet microlensing survey fields  \citep{WFIRST_AFTA}
field, so understanding the source distance distribution in this area of the bulge is
important for WFIRST.        
  
\begin{figure}
\epsscale{1.0}
\plotone{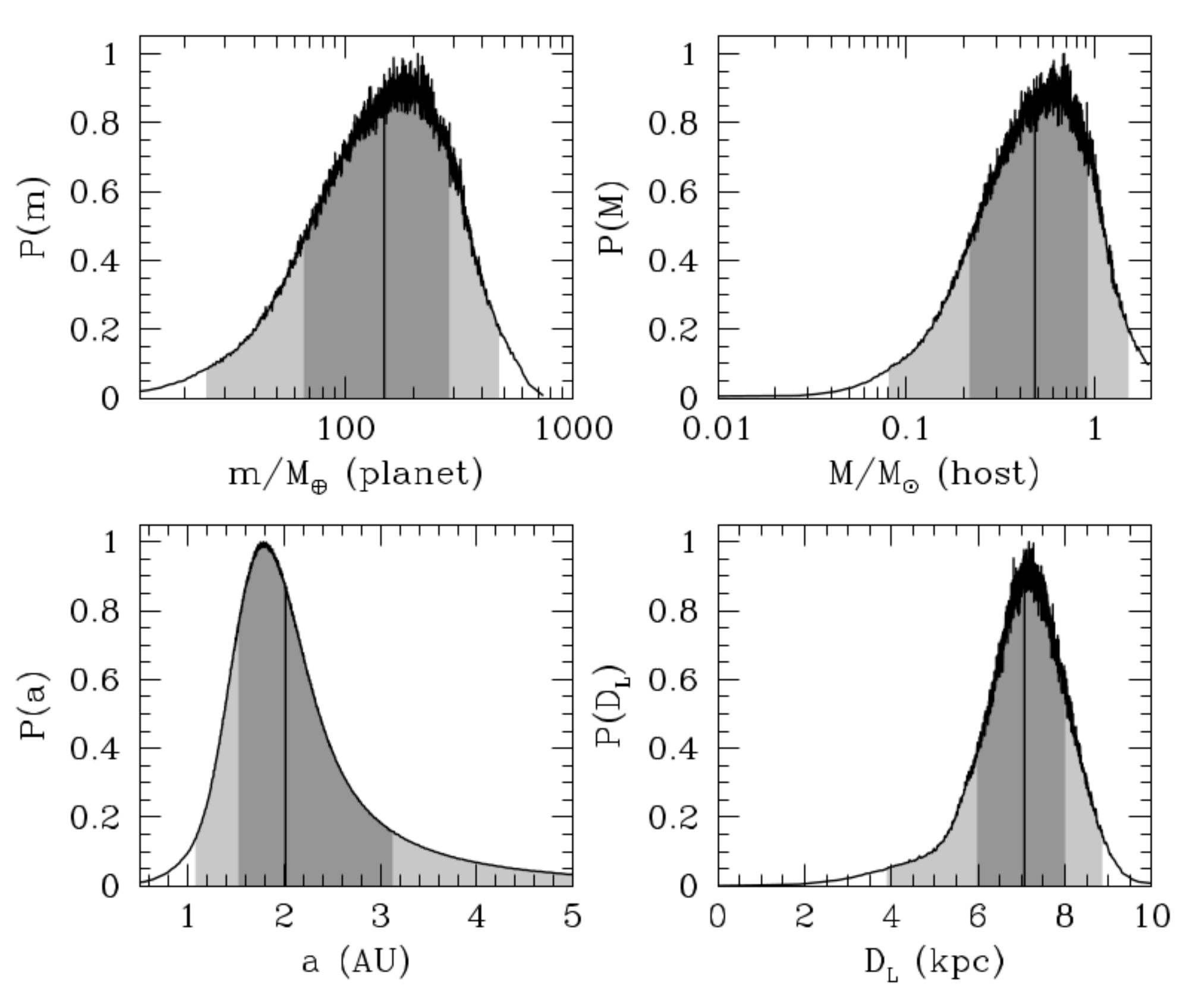}
\caption{The probability distribution of planet mass, host star mass, planet - host star 
separation and distance to lens system,  derived using Bayesian statistical analysis. 
The central $68.3\%$ of the distribution is shaded dark grey and the remaining central
$95.4\%$ of the distribution is shaded light grey. The vertical black line marks the median 
of the probability distribution of respective parameters.}
\label{fig-histogram}
\end{figure}

\begin{figure}
\epsscale{1.0}
\plotone{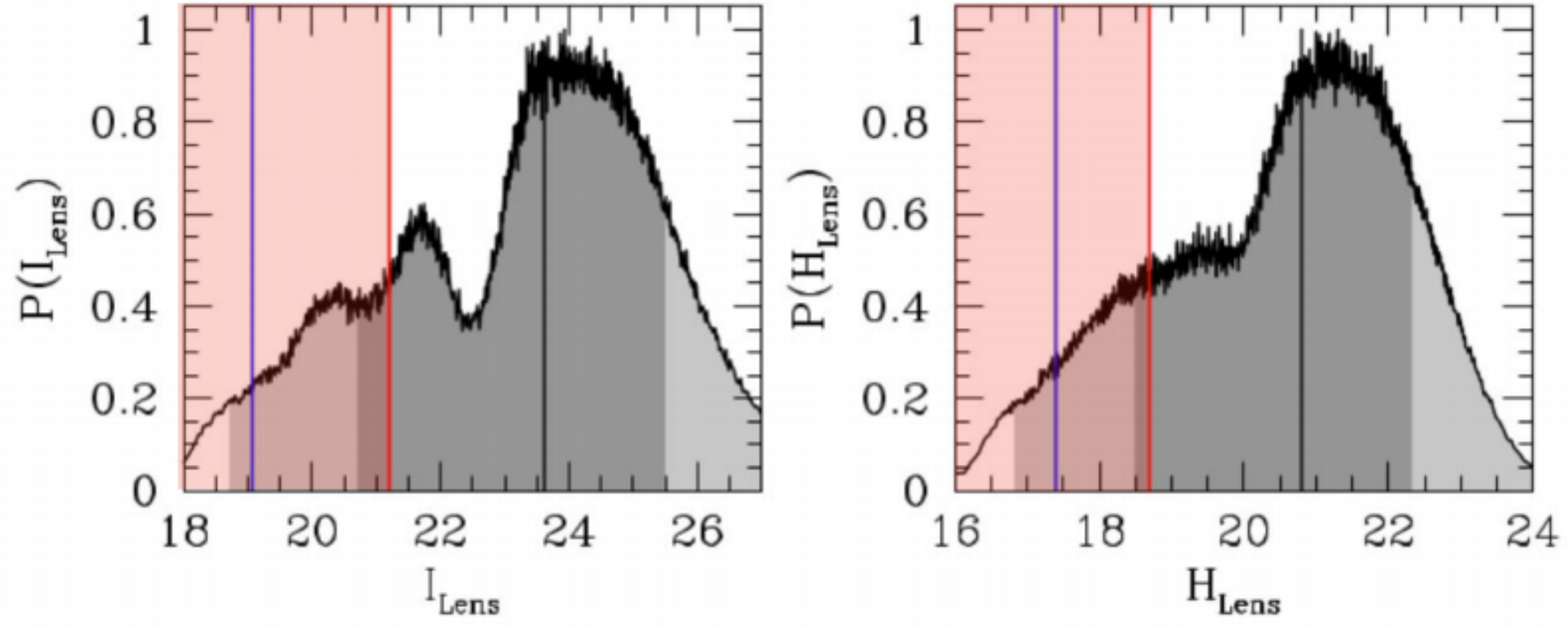}
\caption{The probability distribution for the $I$ and $H$-band lens star magnitudes given the source
radius estimate based on the source magnitude and color given in Figure \ref{fig-cmd}. As in
Figure~\ref{fig-histogram}, central $68.3\%$ is shaded dark grey and the remaining central
$95.4\%$ of the distribution is shaded light grey. The blue lines indicate the source 
magnitude, and the left and right panels show the $I$ and $H$ magnitude distributions, 
respectively. The vertical red lines mark the lens brightness corresponding to the 3-$\sigma$
upper limit on the source brightness, so if the lens is in the light red shaded region it will be
bright enough to push the combined lens+source brightness above the brightness from the 
light curve models. In this case, the excess brightness from the lens will be detectable even 
when the lens and source remain unresolved. If the lens is fainter and located to the right
of the red shaded region, we will not be able to identify the lens with high angular resolution
JWST, HST or adaptive optics observations until the lens star begins to separate from the source
5-8 years after peak magnification.}  
\label{fig-histogram_fmag1}
\end{figure}

The probability distribution for the brightness of the lens star in $I$ and $H$-bands 
is shown in Figure 6. The blend magnitude from best fit is found $I_{bl} = 17.94$ which is 
brighter than the lens star magnitude predicted in Table \ref{tab-params-histogram} and 
Figure 6. Since the lens-source relative proper motion is $\mu_{\rm rel} = 6.55 \pm 1.12\,{\rm mas/yr}$, 
the source and lens will be separated by $\sim 1$ HST pixel ($ 39.6 \pm 6.2$) by 2020. 
If the lens system is observed directly in high resolution images, then the brightness 
measurement of lens leads to the precise calculation of the mass and distance to the 
lens system \citep{bennett06,ogle169, batistaogle169}. But in the $I$-band, the  
source star is brighter than the median prediction for the lens by $\sim 4$ magnitudes 
(see Figure \ref{fig-histogram_fmag1}), which makes it difficult to observe the lens in $I$ band 
with the high resolution follow-up images unless the lens and source are completely resolved. 
The extinction corrected $(I-H)$ color of the source is obtained from $(V-I)$ color using 
\citet{kenyon_hartmann}. The $H$-band extinction $A_{H}$ is also obtained from $A_{I}$ 
and $A_{V}$ using the \citet{cardelli} formula. From the $I$-band magnitude, the $(I-H)$ color and the
$H$-band extinction, $A_{H}$, the $H$-band magnitude of the source is calculated to be 
$H_S = 17.96$. From Figure 6, we see that the $H$-band magnitudes of source and median
prediction for the lens differ by $\sim2.0$-2.5 magnitude. The uncertainty in the source brightness 
is measured from the Markov chain links. We also include the uncertainty in the source color and 
magnitude. There is $5\%$ uncertainty in the color conversion from the extinction corrected 
$(V-I)$ to $(I-H)$. The red lines in Figure 6 is such that if lens is at least as bright as the 
red lines (\ie it lies in the red-shaded region) then lens+source brightness is brighter than the
3-$\sigma$ upper limit on the source brightness. This would mean that the lens could be 
detected when it is still unresolved from the source. Since most of the lens flux histogram 
lies to the right of the red-shaded region, the lens is likely to be too faint to be detected 
in high resolution images unless lens and source are partially resolved. Since the relative 
lens-source proper motion, $\mu_{\rm rel}\sim 6.5\,$mas/yr, the lens and source are 
expected to be resolvable by JWST, HST or adaptive optics imaging some in 6-8 years after
peak magnification, in 2020 - 2022. 
  
\citet{suzuki2016} derived a mass ratio function from planets detected in MOA-II survey data,
and discovered a break in the mass ratio function at $q \sim 10^{-4}$. The 
OGLE-2014-BLG-1760Lb planetary mass ratio of $q = 8.6\pm 0.9$ is above the mass ratio 
break $q_{br}\sim 10^{-4}$. This is close to Jupiter's mass ratio, but the host star probably
has a mass of $M_* \simlt \msun$, so this planet is probably a low-mass gas giant, like
Saturn. Follow up observations with JWST, HST or adaptive optics in 2020-2022 
should be able to measure the lens brightness and determine the planetary mass and 
distance using the methods of \citet{bennett06,bennett07,ogle169} and \citet{batistaogle169}. 
Later with WFIRST \citep{WFIRST_AFTA}, a similar kind of study will provide the
statistics needed to determine the planetary mass function as a function of the host 
star mass and distance.  
  
A.B., D.P.B. and D.S. were supported by NASA through grants NASA-NNX12AF54G and 
NNX13AF64G. I.A.B. and P.Y. were supported by the Marsden Fund of Royal Society of 
New Zealand, contract no. MAU1104. N.J.R. was supported through Royal Society of 
New Zealand Rutherford Discovery Fellowship. A.S, M.L. and M.D. were supported through 
Royal Society of New Zealand. T.S. received support from JSPS23103002, JSPS24253004 and 
JSPS26247023. MOA project received grants from JSPS25103508 and 23340064.  A.F. was supported by the Astrobiology Project of the Center for Novel Science Initiatives (CNSI), National Institutes of Natural Sciences (NINS) (Grant Number AB261005). The OGLE project received funding from National Science Centre, Poland, grant MAESTRO 2014/14/A/ST9/00121 to AU. OGLE team thanks Profs. M. Kubiak and G. Pietrzy\'nski, former members of the OGLE team, for their contribution to the collection of the OGLE photometric data over the past years. DMB was supported by NPRP grant X-019-1-006 from the Qatar National Research Fund (a member of Qatar Foundation). This work makes use of observations from the LCOGT network, which includes three SUPAscopes owned by the University of St Andrews. The RoboNet programme is an LCOGT Key Project using time allocations from the University of St Andrews, LCOGT and the University of Heidelberg together with time on the Liverpool Telescope through the Science and Technology Facilities Council (STFC), UK. This research has made use of the LCOGT Archive, which is operated by the California Institute of Technology, under contract with the Las Cumbres Observatory.  We thank Prof. Andrew Gould and $\mu$FUN team for allowing us to use their data and acknowledge their hard work and contribution in collecting the $\mu$FUN data.  \\\\

\end{document}